\documentclass[12pt]{iopart}

\usepackage{graphicx}
\usepackage{tabularx}
\usepackage{algorithm2e}
\usepackage{multirow}

\begin{document}
\title[Globally optimised model of the cerebral arteries]{Development of a globally optimised model of the cerebral arteries}

\date{\today}

\author{Jonathan Keelan$^1$, Emma M. L. Chung$^{2,3}$ and James P. Hague$^{1}$}
\date{\today}

\address{$^1$School of Physical Sciences, The Open University, MK7 6AA, UK}
\address{$^2$Cerebral Haemodynamics in Ageing and Stroke Medicine (CHiASM) group, Department of Cardiovascular Sciences, University of Leicester, LE1 7RH, UK}
\address{$^3$Medical Physics, University Hospitals of Leicester NHS Trust, Leicester Royal Infirmary, LE1 5WW, UK}

\ead{Jim.Hague@open.ac.uk}

\begin{abstract}
The cerebral arteries are difficult to reproduce from first
principles, featuring interwoven territories, and intricate layers of
grey and white matter with differing metabolic demand. The aim of this
study was to identify the ideal configuration of arteries required to
sustain an entire brain hemisphere based on minimisation of the energy
required to supply the tissue. The 3D distribution of grey and white
matter within a healthy human brain was first segmented from Magnetic
Resonance Images. A novel simulated annealing algorithm was then
applied to determine the optimal configuration of arteries required to
supply brain tissue. The model is validated through comparison of this
ideal, entirely optimised, brain vasculature with the known structure
of real arteries.  This establishes that the human cerebral
vasculature is highly optimised; closely resembling the most energy
efficient arrangement of vessels. In addition to local adherence to
fluid dynamics optimisation principles, the optimised vasculature
reproduces global brain perfusion territories with well defined
boundaries between anterior, middle and posterior regions. This
validated brain vascular model and algorithm can be used for
patient-specific modelling of stroke and cerebral haemodynamics,
identification of sub-optimal conditions associated with vascular
disease, and optimising vascular structures for tissue engineering and
artificial organ design.
\end{abstract}

\maketitle

\section{Introduction}

The brain has an exceptionally high demand for oxygenated blood, and
accounts for 14-20\% of the body's blood supply, despite contributing
only 2\% to body mass \cite{payne2017}. This reflects the
exceptionally high energy requirements of the brain's active grey
matter cells, contained in a thin 2-4 mm layer of cerebral cortex
covering the brain's surface. The brain's requirement for an
uninterrupted supply of blood has resulted in the development of an
exquisite network of arteries. In this study, the first globally
optimised computational model of the cerebral arteries is presented,
developed by minimising the energy required to maintain blood
flow. The most efficient configuration of vessels is identified and
compared with data from human subjects to explore the extent to which
naturally occurring cerebral arterial trees reflect principles of fluid
dynamical optimisation.

Major arteries supplying the brain emerge from the Circle of Willis
(CoW), a ring-like arrangement of arteries positioned at the base of
the brain. The major cerebral arteries (Anterior, Middle, and
Posterior Cerebral Arteries - ACA, MCA and PCA, respectively) then
penetrate the functional part of the tissue (the brain parenchyma) and
branch further to supply the anterior, middle and posterior perfusion
territories. These arteries emerge at the surface (pia) of the brain
to supply the cerebral cortex (see e.g. Payne\cite{payne2017}).  The
cerebral cortex features deep folds (gyri) across its surface, upon
which the pial arteries attach via perforating arteries that perfuse
the grey matter beneath\cite{Cassot2006}. As the vascular tree
branches further into arterioles and capillaries, the tissue
environment becomes more symmetric and homogeneous.  The area of the
vascular bed also increases dramatically, which slows the blood to
allow diffusion of oxygen and capillary exchange. Since oxygen
diffuses slowly but is metabolised rapidly, the majority of brain
tissue lies within (25 microns) of a capillary resulting in shorter,
more numerous, vessels than seen in other organs \cite{payne2017}.

The aim of this study is to identify the most efficient arrangement of
arteries capable of supplying a full brain hemisphere. Idealised
`arteries' in the model vasculature begin at the level of the major
cerebral arteries (radii $\sim$1.5 mm) and branch until reaching
arterioles supplying the capillary mesh $\sim$100$ \mu$m.  Developing
an optimised model of the brain circulation presents a number of
technical challenges; firstly, there are a huge number of vessels with
diameters varying across several orders of magnitude, secondly, the
brain has a complicated geometry with multiple perfusion territories,
finally, the grey and white matter have differing energy
requirements. To identify the most efficient `idealised' arterial
structure capable of supplying an entire brain hemisphere, a simulated
annealing (SA) algorithm is implemented for model optimisation. The
assumption that configurations of vessels in the adult brain have
evolved to be near optimal has not yet been validated through
comparison of an ideal (globally optimally efficient) arterial tree
with real measurements. This methodological milestone is one of the
aims of this study. The ability to design efficient arterial trees has
potential applications in surgical planning, computational modelling
of cerebral haemodynamics, early identification of vascular disease,
improved image segmentation from MR and CT angiography, and the design
of an optimised vasculature to supply artificial tissues or organs.

To determine the most efficient arrangement of arteries, it is
necessary to identify an algorithm capable of minimising the energy
required to supply blood to the tissue, whilst respecting the
functionality of the organ. Murray\cite{Murray1926a} showed that the
sizes of parent and daughter vessels in single optimal bifurcations
would follow the relation
$r_{p}^{\gamma}=r_{d1}^{\gamma}+r_{d2}^{\gamma}$, where $\gamma$ is a
bifurcation exponent describing the relationship between parent and
daughter vessel radii; $r$ the radius of the vessel and subscript $p$
and $d$ represent parent and daughter vessels, respectively. Detailed
topological examination of individual bifurcations in, e.g. the
cerebral vasculature, supports this
relation\cite{rossitti1993}. Deviations from the optimal conditions
predicted by Murray’s law have been shown to be associated with
vascular disease \cite{murraylawdeviations}.

To computationally obtain globally optimised configurations of
arteries, the authors recently proposed a novel Simulated Annealing
Vascular Optimisation algorithm (SAVO) \cite{keelan2016}. Advantages
of this algorithm include the ability to identify the most efficient
`ideal' tree with minimal metabolic demand.  Constraints can be
applied easily during the simulation of e.g. hollow organs. All length
scales are treated using identical optimisation principles. The
algorithm was previously applied to modelling the coronary
vasculature, which involved optimising an extensive ($>$6000 branch)
arterial tree for a challenging hollow organ (the
heart)\cite{keelan2016}. The resulting idealised coronary artery
structure was a close match to porcine cardiac morphological data

In this paper the SAVO algorithm is applied to the cerebral
vasculature. The paper is organised as follows. In section
\ref{sec:method} the algorithm is described, and details of the
segmentation of brain MRI data into gray and white matter are
provided. In section \ref{sec:results}, the resulting arterial trees
are presented and subjected to comparisons with existing morphological
\textit{in vivo} data\cite{data,data2}. A summary and outlook are
presented in section \ref{sec:summary}.

\section{Materials and methods}
\label{sec:method}

In this section, the algorithm used to grow cerebral
arterial trees \textit{in-silico} is detailed. The algorithm is similar to the
approach for growing cardiac vasculature\cite{keelan2016}, with some
differences for using MRI data to provide tissue information, and some
subtleties relating to the supply of cerebral tissue. More
detail is included than in Ref. \cite{keelan2016}.

\subsection{Cost Function}

At the core of the algorithm is a cost function which measures the
fitness of a given tree. It is the sum of (a) the metabolic cost to
maintain blood (b) the cost for pumping blood (c) a requirement to supply
blood evenly to all tissue and (d) a penalty for large vessels that cross
parenchyma.
\begin{equation}
C_{\mathrm{T}} = A_{w,v}(C_w + C_v) + A_o C_o + A_s C_s \label{eq:costfunction}
\end{equation}
where $A_{w,v}, A_{o}$ and $A_{s}$ are dimensionless constants that
scale each contribution to the cost function. $C_v$ is a metabolic
cost for maintaining blood volume and $C_{w}$ is the cost of pumping
through a vessel. In addition, $C_{s}$ is a penalty for over- or
under-supplying tissue and $C_{o}$ is a penalty associated with
vessels that penetrate tissue. To form physiological trees, $A_{s}$,
has a large value since tissue without supply would die. A high
(but slightly lower) value of $A_o$ heavily penalises vessels that cut
through parenchyma to affect organ function. Therefore $A_s$ and $A_o$
act as constraints.

\subsubsection{Pumping cost}

Following Murray\cite{Murray1926a}, Poiseuille flow is assumed to
calculate the power dissipated during flow through a vessel, $W_i$,
\begin{equation}
W_i = \frac{8\mu l_i f_i^2}{\pi r_i^4}
\end{equation}
where $f$ is volumetric flow rate, vessels are cylindrical with radius
$r$ and length $l$ and a representative value is used for the viscosity
of blood $\mu=3.6\times 10^{-3}$Pa s (although it is noted that blood
viscosity can drop significantly in small vessels $<~
100\mu$m\cite{fahraeus1931}). The total power dissipation, $C_w
=\sum_{i}W_{i}$ is calculated by summing over all vessels, $i$.

For computational efficiency, a constant input flow is maintained, and
terminal flows are fixed and equal. Thus, from conservation of mass,
the flow in any artery depends only on the number of terminal sites
downstream. If $f_{\rm root}$ represents total flow into the tree, the
total flow per end node is $f_{\rm term}=f_{\rm root}/N_{\rm end}$. A
power law is used to relate radii and flow\cite{west1997general}, $f =
\epsilon r^{\gamma}$, where $\gamma$ is the bifurcation exponent which
is set from experimental considerations and $\epsilon$ is a constant
determined from $f_{\rm root}$ and $r_{\rm root}$.

\subsubsection{Metabolic cost for maintenance of blood volume}

Following Murray\cite{Murray1926a}, a metabolic cost to
maintain a volume of blood is assumed:
\begin{equation}
C_v = m_b \sum_{i} V_{i}.
\end{equation}
The value $m_b=648$J s$^{-1}$ m$^{-3}$ is chosen, which is within the
measured range for humans\cite{Liu2007}. $V_{i}$ is the total volume
of a cylindrical arterial segment.

\subsubsection{Blood Supply Penalty}

In a healthy organism, perfusion is even throughout tissue over a
physiologically long timescale\cite{homogenousdistribution}. Thus,
terminal nodes need positioning so that the supply is uniform. The
terminal nodes in SAVO are much larger than capillaries, so terminal
nodes are considered to be surrounded by spherical microcirculatory
``black boxes''\cite{Schreiner1993} of radius $R_{\rm supply}$. Within
each sphere the details of the microvasculature are neglected. Without
this assumption the computational power required to optimise the tree
would be prohibitive.  The radius of the spheres is calculated using
physiological values for the blood demand of the tissue as,
\begin{equation}
4\pi R_{\rm supply}^3 / 3 = f_{\rm term}/q_{\rm req},
\label{eqn:rc} 
\end{equation}
where $q_{\rm req}$ is the volumetric blood flow required to maintain
tissue. Gray and white matter in the brain have different supply
requirements\cite{cbfrelval}, $q_{\rm req, gray} = 10.9\times
10^{-3}\mathrm{s}^{-1}$ and $q_{\rm req, white} = 3.57\times
10^{-3}\mathrm{s}^{-1}$ respectively, leading to a smaller $R_{\rm
  supply}$ and thus higher density of terminal nodes sited in the gray
matter. The total flow is determined using $q_{\rm req}$
and the total volume of gray and white matter in the MRI scans, which
are $389.12 \times 10^{-6}\mathrm{m}^3$ and $321.64 \times
10^{-6}\mathrm{m}^3$ respectively.

There is no unique way to define the supply penalty. For this paper,
\begin{equation}
C_s = \sum_{\mathbf{voxels}} s;\, 
s = \left\{ \begin{array}{rl}
 10 &\mbox{ if $b = 0$} \\
  (b - 1)^2 &\mbox{ otherwise}
       \end{array} \right.
\end{equation}
where $b$ is the total number of spheres contributing to the supply of
a single tissue voxel and the sum is performed over all tissue
voxels. $C_s$ favours voxels supplied by a single sphere, and thus
encourages dense packing of supply spheres while minimising
overlap. This cost only needs to be recalculated when terminal nodes
move.

\subsubsection{Exclusion of Large Arteries}

Larger arteries are absent from regions of many organs as they would
interfere with function. For example the brain has pial arteries
running across its surface, whereas only the smaller branches from
these arteries are allowed to penetrate. Also, the higher blood supply
requirements of grey matter\cite{cbfrelval} may play a role in
bringing larger arteries to the surface.

A distance map is used to calculate a penalty that increases with
vessel depth within the excluded tissue, thus enforcing exclusion of
larger vessels. A cutoff radius $R_{\rm ex}$ is defined whereby any node
exceeding this radius incurs a cost,
\begin{equation}
C_o = \sum_{R > R_{\rm ex}} D_{ijk\in S}^6,
\end{equation}
where $i$, $j$ and $k$ are voxel coordinates, $S$ is the line segment
corresponding to the vessel and $D_{ijk}$ is the distance map value,
which is calculated from MRI data as discussed in the next section
\cite{keelan2016}.

Cerebral arterioles with diameters less than $\sim100 \mu m$ are
responsible for penetrating deep within cortical
tissue\cite{Cassot2006}. For even the largest trees considered here
the smallest vessels are slightly wider than this value.  $R_{\rm
  ex}=150\mu$m is selected to ensure some penetration into the tissue,
and to understand how results are modified by changes in $R_{\rm ex}$.

\subsection{Tissue voxel map and MRI data}

\begin{figure}
  \centering
  \includegraphics[width = 0.48\textwidth]{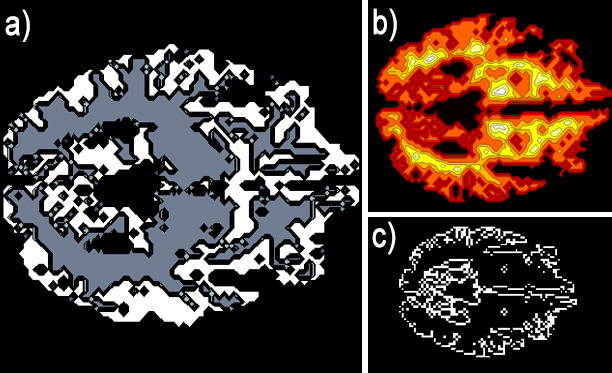}
  \caption{An image slice of the MRI data from a healthy individual,
    at various stages of analysis. Panel a) shows the segmented
    tissue, b) shows the result of the distance map calculation, c)
    shows the identified surface voxels.}
  \label{fig:brain_data}
\end{figure}

In a modification to the previous algorithm\cite{keelan2016}, MRI data
are used to provide realistic tissue shapes. Both a T$_{1}$-weighted
image and time-of-flight (TOF) angiogram from a healthy individual
(one of the authors) were collected using a 3T Siemens Skyra MR
scanner (Siemens Medical, Erlangen, Germany). Spatial resolution for
the two images was 1 $\times$ 1 $\times$ 1 mm$^{3}$ and 0.5 $\times$
0.5 $\times$ 0.5 mm$^{3}$, respectively. Images were exported for
processing in DICOM format. The angiogram was used to locate the root
positions of the MCA, ACA and PCA. The T$_{1}$-weighted image was
segmented into white and gray matter using the statistical parametric
mapping function in MATLAB (MathWorks, Natick, MA, USA)
\cite{spm8}. Before calculation, MRI images were downscaled to reduce
computational time. Surfaces of the combined white and gray matter
were identified via a nearest neighbour search, where voxels having at
least one unoccupied neighbour were labelled as belonging to the outer
surface of the brain. Each voxel in the tissue was assigned a value
equal to the shortest distance to the surface of that space (the
distance map\cite{distancemap}). A sample slice at each stage of the
segmentation process is shown in Fig. \ref{fig:brain_data}. Following
segmentation, the brain is divided into left and right hemispheres and
SAVO is performed for a single hemisphere. This reduces the total number of
nodes in an already challenging arterial tree, and is justified since
vessels downstream of the Circle of Willis do not typically cross
between hemispheres.

Since MCA, PCA and ACA inputs are very closely spaced, a single
arterial inlet is provided to the SAVO algorithm, representing the
approximate position of these arteries as they branch from the CoW.
Relative flows are then calculated by the optimisation
algorithm. Currently it is not possible to predict the structure of
anastamoses such as the CoW, however anastomoses make up a tiny
proportion of all vessels.

\subsection{Simulated Annealing}

There are many options for solving the cost function to obtain a
vasculature. SA is used, which is a general optimisation technique
inspired by the physical process of annealing\cite{siman}. Sequential
random updates, are made to trial solutions. The probability of
accepting each update is given by,
\begin{equation}
P_{ij} = {\rm min}\left\{\exp(\frac{-\Delta C_{ij}}{T}),1\right\}\label{annealing}
\end{equation}
where $P_{ij}$ and $\Delta C_{ij}=C_{j}-C_{i}$ are the probability and
change in cost associated with altering tree configuration $i$ to
configuration $j$ respectively. $T$ is the annealing temperature,
which is slowly reduced. $P_{ij}$ permits occasional acceptance of modifications that
increase the cost, allowing the solution to climb out of local
minima.

Annealing temperature must reduce slowly enough to explore the
configuration space thoroughly. The initial temperature, $T_{\rm init}$ is
chosen to be much higher than the cost change associated with a
typical update and $T_{\rm final}$ much lower. The following temperature
schedule is used,
\begin{equation}
  T_{n + 1} = \alpha T_{n} \label{eq:satempschedule}
\end{equation}
where $n$ denotes the SA step and 
\begin{equation}
\alpha = e^{\frac{1}{S}\left( \ln T_{\mathrm{init}} - \ln T_{\mathrm{final}} \right)}.
\end{equation}
where $S$ is the total number of SA steps.

\subsection{Arterial tree updates}
\label{subsection:moves}

Arteries are represented as a bifurcating tree. The root node
represents the largest artery. The tree branches at bifurcation
nodes. The simulated tree is truncated with $N_{\rm end}$ terminal
nodes, representing the smallest vessels that can be modelled.  Each
bifurcation and terminal node of the tree has a 3D
coordinate. Bifurcations are connected via straight vessels.  Due to
computational constraints, terminal nodes are typically larger than
the arterioles directly feeding capillaries in living organisms.

At each SA step, either of the following updates are made: (1) move a
node, or, (2) swap node connections to change the tree topology. This
minimal set of updates guarantees ergodicity (i.e. the algorithm can
explore any tree configuration). For update (1):
\begin{enumerate}
  \item Randomly choose a bifurcation from the tree.
  \item For each spatial dimension generate a uniform random number between $-d_{\mathrm{move}}$ and $d_{\mathrm{move}}$, where $d_{\mathrm{move}}$ is the maximum displacement distance.
  \item Add the displacement to the randomly chosen bifurcation.
\end{enumerate}
In practice, the tree optimises more quickly using two different values of $d_{\mathrm{move}}$.

For update (2):
\begin{enumerate}
  \item Randomly choose two nodes.
  \item If either node is upstream of the other, a new pair of nodes
    is selected randomly until a valid pair is found
  \item Swap the parents of each node.
\end{enumerate}
Updates are summarised in table \ref{table:moveweights}. The root node is never updated.

\begin{table}
  \renewcommand{\arraystretch}{1.5}
  \centering
  \caption{Weightings, parameters, and target nodes for updates in the SA algorithm. Update are selected with the probability weight shown.}\label{table:moveweights}
  \begin{tabular}{|c | c | c | c | c |}
    \hline
    Type & Parameter & Target  & Weight \\
    \hline
    \multirow{4}{*}{Move} & $d_{\mathrm{move}} = 1$mm & bifurcation & 0.675 \\
     & $d_{\mathrm{move}} = 1$cm & bifurcation & 0.075 \\
     & $d_{\mathrm{move}} = 1$mm & terminal node & 0.045 \\
     & $d_{\mathrm{move}} = 1$cm & terminal node & 0.005 \\
    Swap  & none & any node & 0.2 \\
    \hline
  \end{tabular}
\end{table}

\subsection{Outline of the algorithm}

The whole SAVO algorithm is now summarised. The initialisation steps
are as follows:
\begin{enumerate}
\item Generate a maximally asymmetric tree with $N=2N_{\rm end}-1$ terminal nodes by
  (a) connecting the root node to a terminal node (b) introducing a
  connection node between the terminal node and its parent, and
  attaching another terminal node to this connection (c) repeating
  steps (a) and (b) $N - 1$ times.
  \item Assign the position of each non-terminal node of the tree to a
    random location in space.
  \item Assign each terminal node a random position inside the tissue
    area to be perfused.
  \item Randomise the topology of the tree by repeatedly applying the
    swap node move 1000 $\times$ \textit{N} times.
  \item Traverse through the tree from a terminal node to the root,
    add $f_{\rm term}$ to the flow of each node visited. Repeat for
    all terminal nodes.
  \item For each node, calculate its radius $(f/\epsilon)^{1/\gamma}$.
\item Calculate the initial value of the cost function using
  Eq. \ref{eq:costfunction}.
\end{enumerate}
Once initialised, the tree is in a random but valid topological and spatial state. The optimisation procedure is as follows:
\begin{enumerate}
  \item Randomly choose and apply an update using the weightings found in table \ref{table:moveweights}.
  \item Calculate the cost function (Eq. \ref{eq:costfunction}) for the newly modified tree.
  \item Using Eq. \ref{annealing}, accept or reject the modification. If rejected, revert the tree to its previous state.
  \item If the cost function of the trial vasculature is smaller than all previous states, then record the tree state.
  \item Reduce the SA temperature following Eq. \ref{eq:satempschedule}.
  \item Repeat the previous steps $S$ times.
\end{enumerate}

A full list of parameters used in the SAVO algorithm is provided in
table \ref{table:cerebralparams}. For large trees, the total number of
topological states is vast, and good optimisation can only be achieved
for very large numbers of SA steps.
The optimisation of the 8191 node tree
in a realistic geometry takes several days on an Intel i7 2.8GHZ desktop
PC. This timescale grows rapidly with tree size. The Open University
IMPACT cluster was used to carry out multiple calculations with
different random number seed (RNS) and large tree sizes. Several
anneal runs were made for each parameter set with different RNS. For
the displayed data the variance on the final cost function from these
runs was small, indicating results from the algorithm are near
optimal. The tree with the smallest cost function is analysed.
\begin{table*}
  \centering
 \caption{Input parameter names, symbols, values and typical sources.}\label{table:saparams}\label{table:weightings}\label{table:cerebralparams}\label{table:inputparams}
  \renewcommand{\arraystretch}{1.5}
  \begin{tabular}{|c|c|c|c|}
    \hline
    Parameter & Symbol & Value & Source \\
    \hline
Flow requirement (gray) & $q_{\rm rec,gray}$ & $10.9\times 10^{-3}$ (m$^3$/s)/m$^3$ brain & SPECT\\
Flow requirement (white) & $q_{\rm rec,white}$ & $3.57\times 10^{-3}$ (m$^3$/s)/m$^3$ brain & SPECT\\
Volume (gray) & $V_{\rm gray}$ & $389.12 \times 10^{-6}$ m$^{3}$ &  MRI \\
Volume (white) & $V_{\rm white}$ & $321.64 \times 10^{-6}$ m$^{3}$ &  MRI \\
Root flow & $q_{\rm rec,gray}V_{\rm gray}$ & &  \\
& $+  q_{\rm rec,white}V_{\rm white}$ & $323.4$ ml / min & Calculated \\
    Root radius & $r_{\mathrm{root}}$ & 1.5 mm & Physiology \\
    Root position & N/A & Average of MCA, PCA, ACA & TOF MRI \\
    Branching exponent & $\gamma$ & 3.2 & MRI \\
    Metabolic constant & $m_b$ & 648 J s$^{-1}$m$^{-3}$ & PET \\
Node exclusion parameter & $R_{\rm ex}$ & 150$\mu$m & Physiology \\
No. end nodes & $N_{\rm end}$ & 4096 & Selected \\
    SA steps & $S$ & $10^{10}$ & Selected \\
    SA initial temperature & $T_{\mathrm{init}}$ & $10^{12}$ & Selected \\
    SA final temperature & $T_{\mathrm{final}}$ & $10^{-10}$ & Selected \\
    Cost function weight & $A_{w,v}$ & $1 \times 10^{4}$ & Selected \\
    Exclusion penalty & $A_o$ & $1 \times 10^{15}$ & Selected \\
    Supply penalty & $A_s$ & $1 \times 10^{30}$ & Selected \\    \hline
  \end{tabular}
\end{table*}

\subsection{Comparison data}

Comparison trees were taken from the BraVa
database\cite{data,data2}. Wright \textit{et. al.} extracted cerebral
arterial tree morphometry from 3T time-of-flight MRA high-resolution
images of 61 healthy volunteers, and then segmented the trees manually
from MRI image slices using the ImageJ software package. The data were
reanalysed to use diameter defined Strahler order (DDSO) which leads
to better classification of vessel segments\cite{Jiang1985}. A
modified DDSO (MDDSO) procedure was used to account for data that are not
Gaussian distributed (see appendix). Trees were discarded from the
dataset if they had any arteries labelled as zero radius or if the MDDSO
algorithm failed to converge. The majority of trees from human
subjects had 5 MDDSOs, and trees with 5 and 6 MDDSOs were analysed
separately. The analysis procedure was identical for experimental and
computational trees.

\begin{figure}
  \centering
\includegraphics[width=70mm]{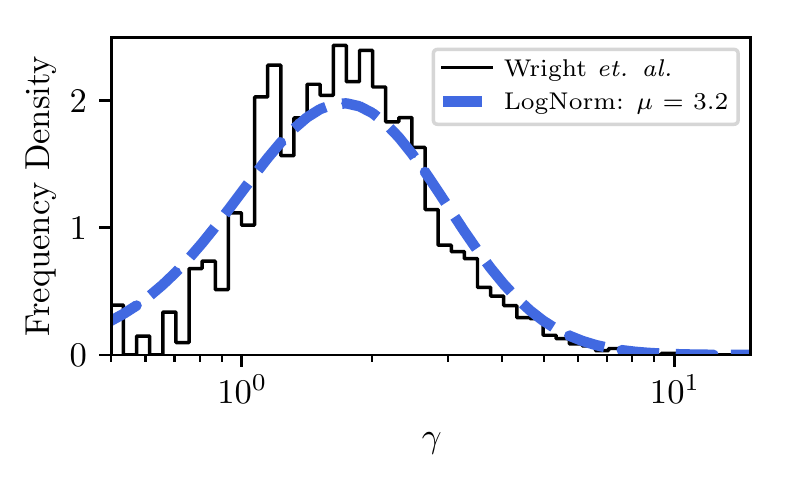}
\caption{Histogram showing the distribution of $\gamma$ values in the
  BraVa data\cite{data,data2}. Fit of a log
  normal curve to the histogram finds a mean value, $\mu$, of 3.2 for
  $\gamma$.}
\label{fig:bifexp}
\end{figure}

Sensitivity analysis for 2D trees (to be published separately)
indicates that the tree structure is only sensitive to the bifurcation
exponent, $\gamma$. Bifurcations in the BraVa data were analysed to
estimate the bifurcation exponent. Results are shown in
Fig. \ref{fig:bifexp}, with mean value found to be
$\gamma=3.2$. Therefore, $\gamma=3.2$ was used for all computational
trees.

\section{Results}
\label{sec:results}

\begin{figure}
  \centering
\includegraphics[width = 0.6\textwidth]{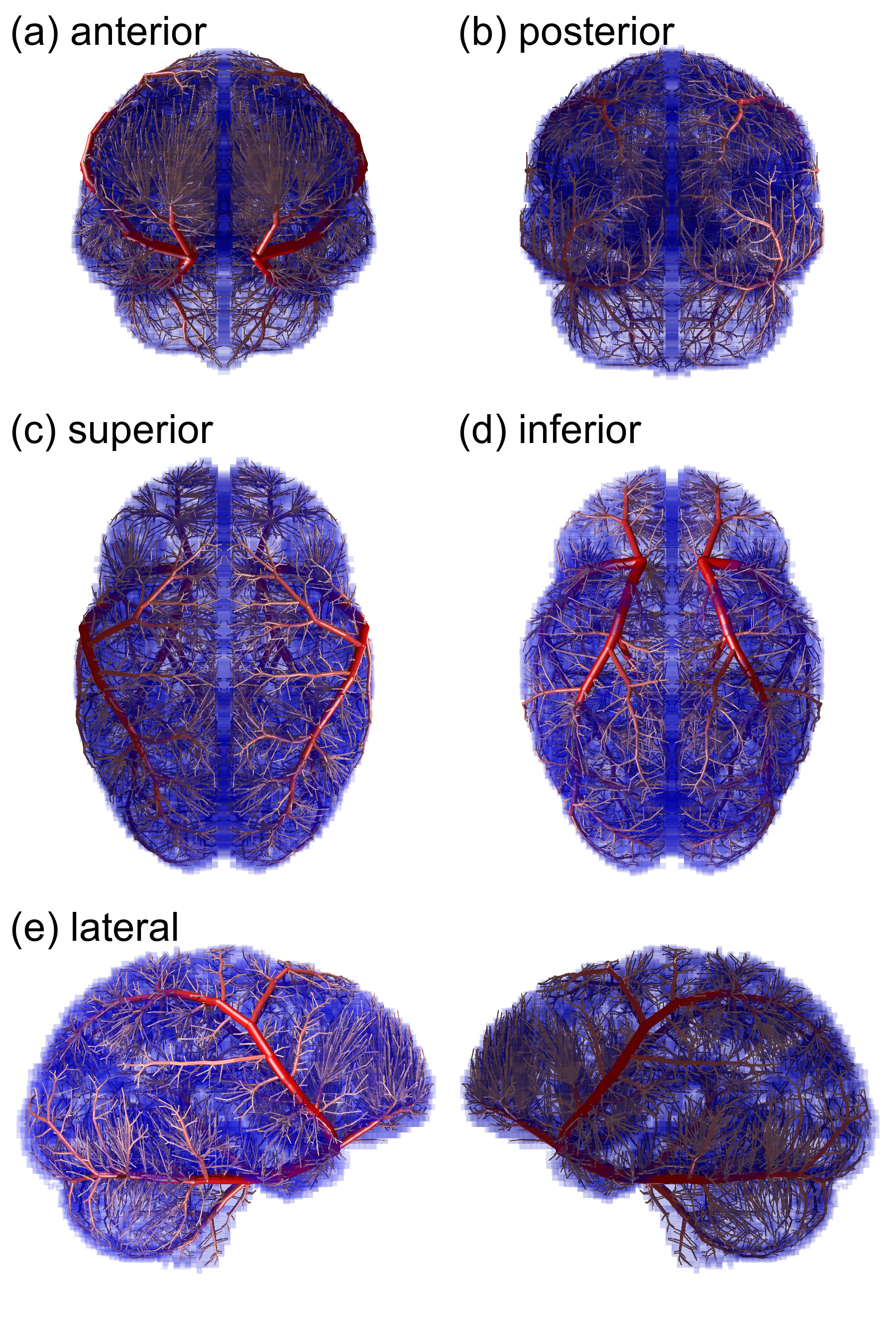}
  \caption{Cerebral vasculature automatically generated on a geometry
    obtained from MRI imaging (8191 seg.). The vasculature is grown on
    a single hemisphere and then symmetrised. Vessel radius has been
    doubled in the images to improve visibility.}
  \label{fig:brain_outside}
\end{figure}

Figure \ref{fig:brain_outside} shows the appearance of the vasculature
as viewed from various angles generated for a single hemisphere with
$N=8181$, and symmetrised about the centre of the brain before
rendering with POV-ray (public domain). The exclusion radius $R_{\rm
  ex}=150\mu$m, input radius is 1.5mm. Large sections of the arteries
run across the outer surface of the brain, mimicking the pial
arteries. While the computed arterial trees lack tortuosity, similar
forms of the major arteries are seen. Looking from the bottom view,
the single input quickly divides into 3 large arteries supplying the
front, rear and side of the brain. The large artery to the front
roughly corresponds to the anterior cerebral artery (ACA), to the side
to the middle cerebral artery (MCA) and the large artery directed to
the rear the posterior cerebral artery (PCA) and cerebellar
arteries. The approximate form is very similar to textbook schematics
of the cerebral arteries that can be found in
e.g. Ref. \cite{graysanatomy}. The algorithm presented here only
considers branching trees with no mechanism to generate anastamoses,
so the Circle of Willis is not reproduced.

\begin{figure}
  \centering
\includegraphics[width = 0.6\textwidth]{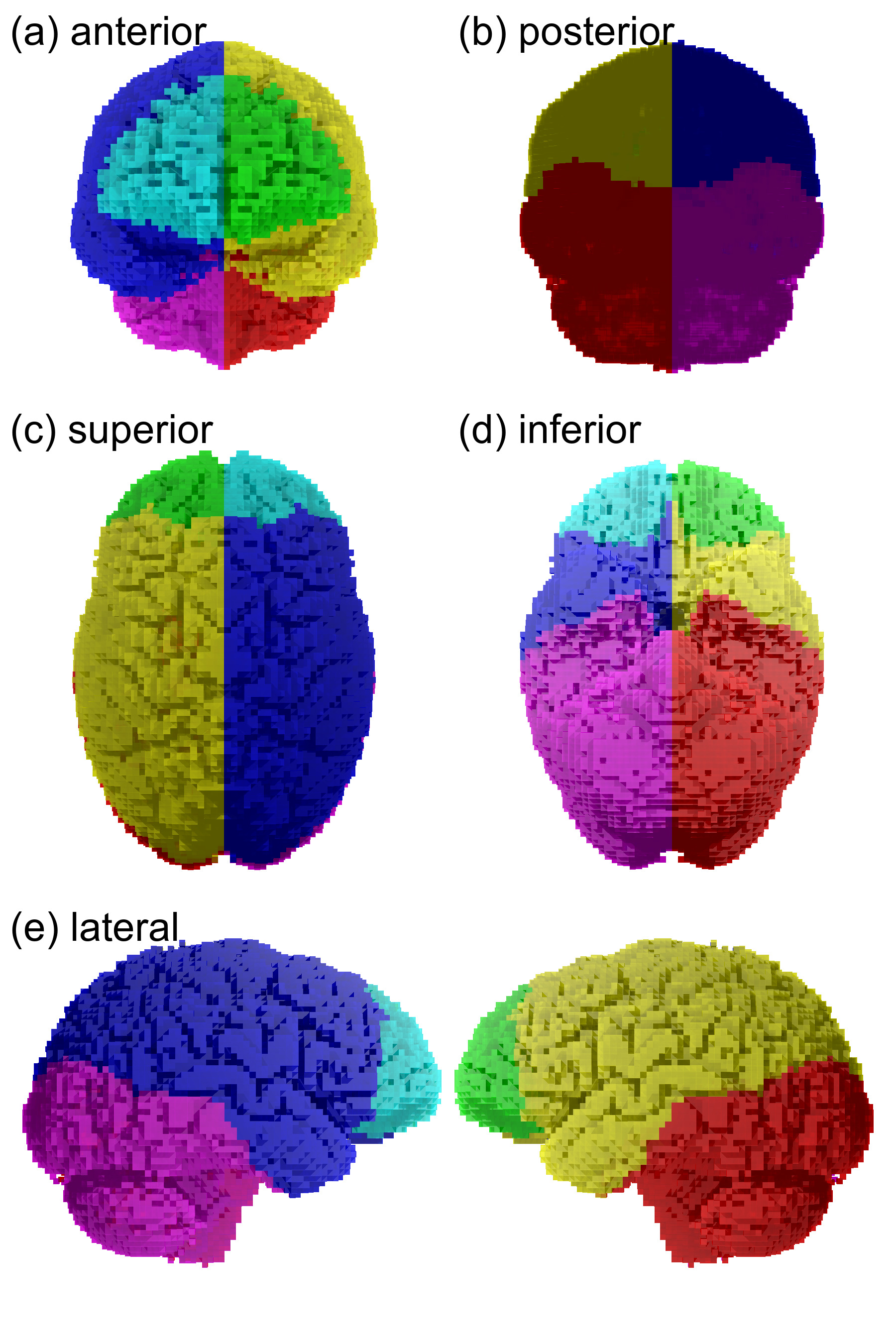}
  \caption{Perfusion territories of the three large generated vessels
    viewed from various angles. Each territory is denoted with a
    different colour. The perfusion territories have well defined
    boundaries between anterior, middle and posterior regions.}
  \label{fig:territories}
\end{figure}

Figure \ref{fig:territories} shows the perfusion territories of the
three large vessels emanating from the input vessel, with each colour
representing a perfusion territory. Again the single hemisphere has
been symmetrised about the central axis.  Views are shown from a
variety of directions.  The perfusion territories are well
differentiated between the anterior, middle and posterior regions of
the brain, consistent with clinical observation \cite{data}. The MCA
territory (yellow and blue) flows up the fissure of Sylvius before
occupying most of the upper region of the brain. The ACA territory
(green and cyan) is found towards the front of the brain. The third
territory (purple and red) supplies the rear of the brain and the
cerebellum.

\begin{figure}
  \centering
\includegraphics[width = 0.48\textwidth]{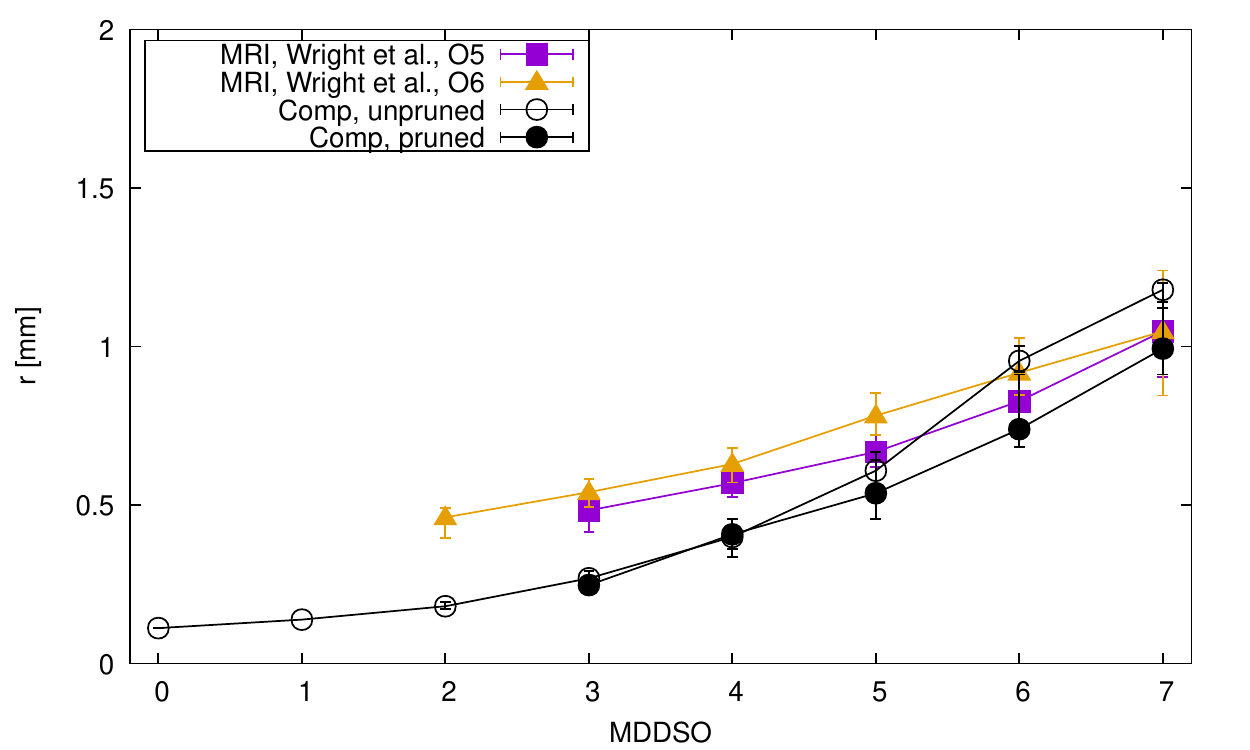}
  \caption{Plot showing the mean radii of the generated tree vs the experimental data of Wright \textit{et. al.}\cite{data,data2} as a function of modified diameter defined Strahler order (MDDSO). Bars show 25th and 75th percentiles.}
  \label{fig:mcaradii}
\label{fig:radhist}
\end{figure}

 The first panel of Figure \ref{fig:mcaradii} shows the mean arterial
 radius, r, vs MDDSO. MRI data were
 averaged over all trees, whereas the computational data is averaged
 over only the tree with lowest cost. Error bars show 25th and 75th
 percentiles of the data. For the radius, general agreement with
 experimental data is good. The levelling out of the radius seen in the
 MRI data for smaller Strahler orders may be related to overestimation
 of the smallest radii in the MRI data due to resolution effects. The
 overall behaviour of the radii as a function of branching order is
 matched between the experimental and generated data.

\begin{figure}
  \centering
\includegraphics[width = 0.48\textwidth]{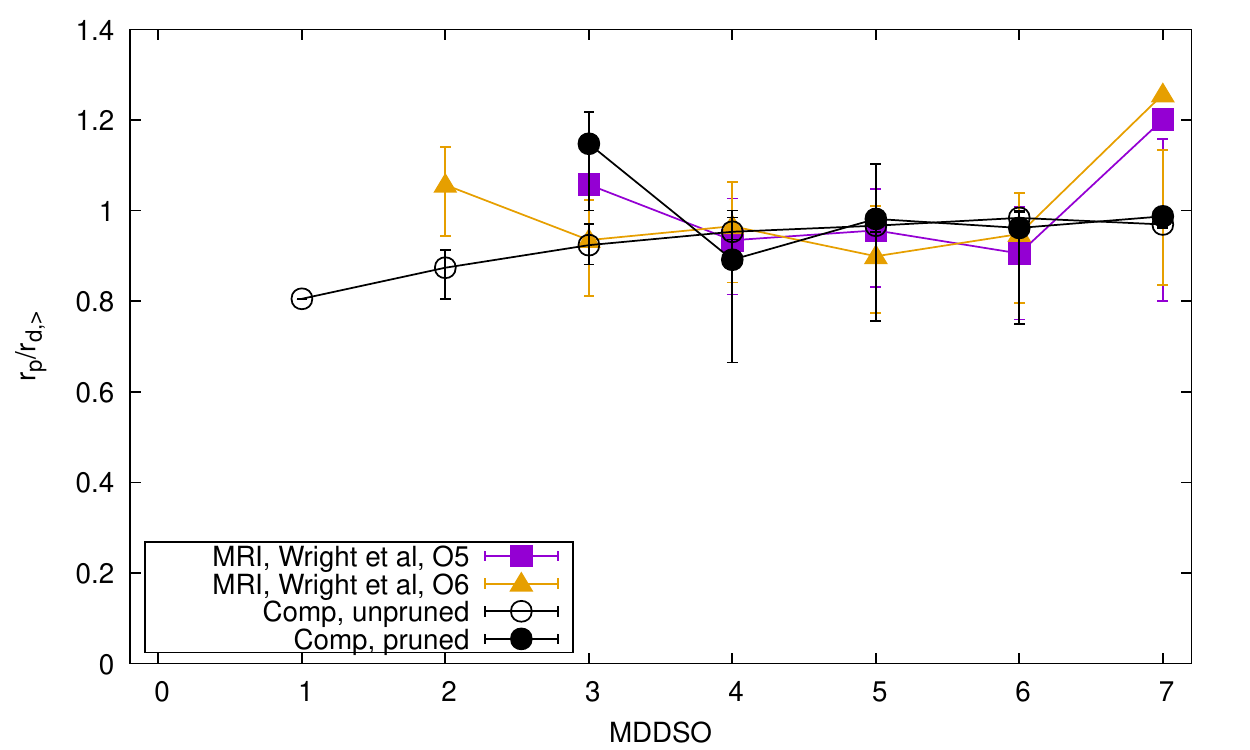}
\includegraphics[width = 0.48\textwidth]{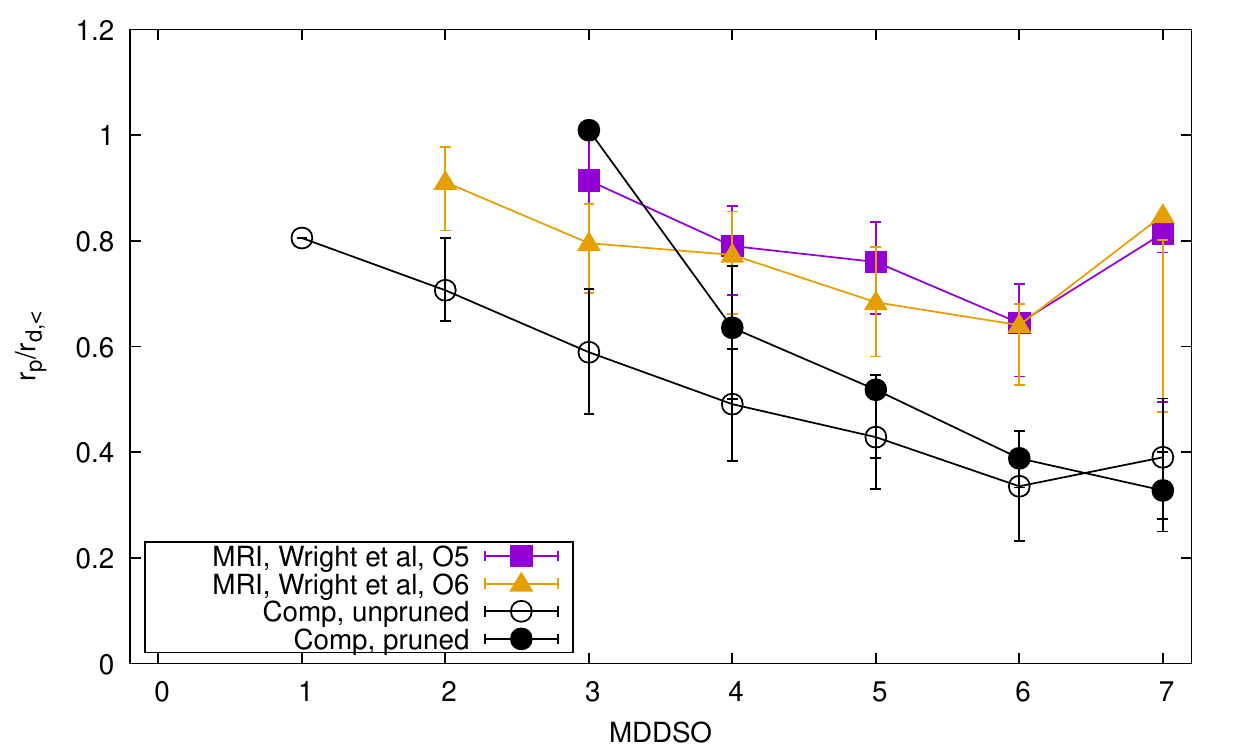}
\caption{Plot showing the asymmetry ratios, $r_{p}/r_{d>}$ and $r_{p}/r_{d<}$, vs MDDSO for MRI and {\it in-silico} data. Bars show 25th and 75th percentiles.}
\label{fig:branching_expt}
\end{figure}

Next, the asymmetry ratios, $r_{p}/r_{d>}$ and $r_{p}/r_{d<}$, are
examined, where $r_{p}$ is the parent vessel radius, $r_{d>}$ and
$r_{d<}$ are the radii of the larger and smaller daughter vessels in
the bifurcation respectively, which are shown in
Fig. \ref{fig:branching_expt}.  Branching ratios are expected to tend
to $1/2^{1/\gamma}$ (approximately 0.8 for $\gamma=3.2$) since
arterial trees must necessarily become more symmetric as they become
smaller. This can be understood by considering the final arterioles
before the capillary bed, which are of roughly equal size, so the
final bifurcation before the capillary bed must be roughly symmetric.
An oddity of the BraVa MRI data \cite{data,data2} is that the
branching ratio of the largest branch is bigger than 1 for lowest
Strahler order, which is surprising because it indicates that some
vessels get wider after branching (rather than smaller as is normally
the case). This effect occurs due to an effective discretisation of
the MRI data due to the 0.31mm voxel resolution. In order to get a
more meaningful comparison of the generated trees to the BraVa
database\cite{data,data2} the effects of MRI resolution are replicated
in the computational trees. The effect of the MRI and subsequent
segmentation is to prune small arteries from the tree and discretise
the radius, which is relatively straightforward to replicate in the
generated trees by including only the arteries with radius greater
than a cutoff, $R_{C}$, and discretising the radius in steps of
$R_{C}$ above this value by rounding down to the nearest multiple of
$R_{C}$ (this discretisaton originates from voxelisation and is seen
in the BraVa data). For around 20\% of the radii, an additional voxel
width is added at random to emulate aliasing effects for vessels that
sit close to voxel boundaries and therefore appear to be wider than
their true width by an extra voxel. $R_{C} = 0.228\mathrm{mm}$ is
chosen so that it is as close as possible to the MRI resolution
cutoff, while maintaining the same number of Strahler orders as the
MRI data.

\begin{figure}
  \centering
\includegraphics[width=0.48\textwidth]{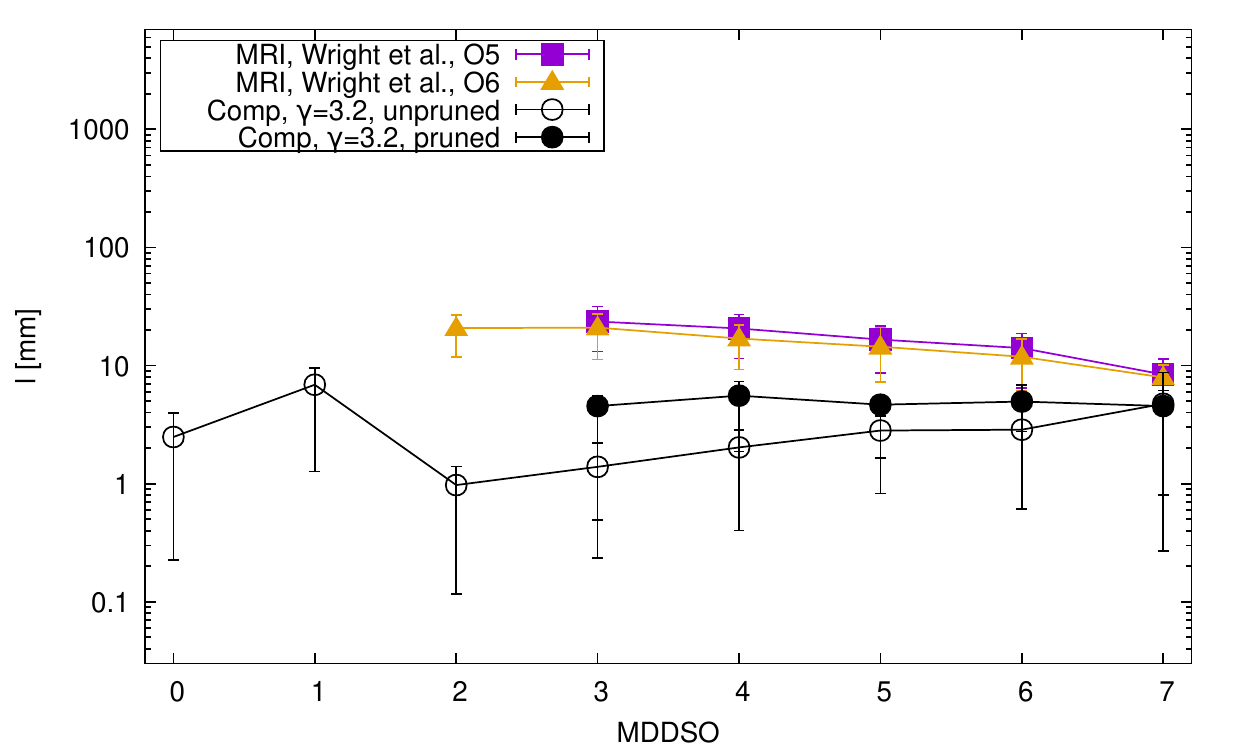}
\caption{Plot showing the mean lengths of various branches of the generated tree in comparison to the experimental BraVa data\cite{data,data2} as a function of branching order. Bars show the 25th and 75th percentiles.}
\label{fig:lengths}
\label{fig:lengthhist}
\end{figure}

Finally in Fig. \ref{fig:lengths} the relationship between length and
branching order for the generated tree is examined. The first panel of
Fig. \ref{fig:lengths} shows the mean lengths of the vessels vs
MDDSO. For length measurements, the pruning procedure is very
important, since the effect of finite MRI resolution is an apparent
lack of small branches from major vessels in the MRI data, potentially
leading to overestimation of segment length. On the other hand, the
computational data have only short lengths between bifurcations due to
large numbers of small vessels branching from major arteries that
would not be imaged by MRI, and thus obscure the comparison. Examining
the pruned data, the mean lengths of the largest Strahler order are
roughly consistent between measured and {\it in-silico} trees. Lengths
of vessels in the MRI data appear to increase slightly as radius
decreases, whereas the pruned computational data has roughly constant
length. It is expected that smaller $R_{\rm ex}$ leads to small increases in
length at the highest Strahler order, because the vessels would have
to follow the surface of the brain over a longer distance before
penetrating the tissue. Ideally, trees would be grown using the
physiological value of $R_{\rm ex}=50\mu$m. It is estimated that
meaningful examination of trees with $R_{\rm ex}=50\mu$m would require
trees with the order of 100000 segments, so that the smallest vessel
radius is less than $R_{\rm ex}$. This is outside our current
computational capability.

\section{Discussion}
\label{sec:summary}

Modelling the vascular structure of the brain presents new
challenges. Unlike the heart, which can be modelled as comprising of
only myocardial tissue, the brain is composed of grey and white matter
with very different metabolic demand. These two tissue types have
differing volumetric blood flow requirements per mass of
tissue\cite{cbfrelval}, which must be factored into the
arterial tree generation algorithm. To the best of our knowledge,
algorithms for arterial growth capable of global optimisation have not
been applied to the vasculature of the brain.

In this article, a simulated annealing based method is applied to {\it
  in-silico} growth of the arterial trees supplying the
brain. Morphological analysis shows that the radii and asymmetry
ratios of vessels in the human brain are well represented by the
optimised trees. The lengths of vessels are shorter than found in MRI
data. Agreement in lengths improves significantly once the pruning
effects of MRI resolution are taken into account. The visual structure
of the generated trees compares well with the form of the major
arteries, especially the pial arteries that traverse the surface of
the brain and MCA traversing the fissure of Sylvius, and three major
arteries corresponding to MCA, PCA and ACA. The morphological
comparison between {\it in-silico} and real arteries indicates that
the structures of the brain have been highly optimised by evolution to
minimise energy consumption.

It is appropriate to put the algorithm presented here into the context
of several other schemes for {\it in-silico} arterial `growth'. Early
approaches use stochastic methods based on morphological
data\cite{wang1992}. A number of algorithms attempt to mimic
`sprouting angiogenesis'. This works best when modelling disordered
arterial trees associated with malignant tumour growth\cite{ref1}.
However, such algorithms are not yet generalisable to the growth of
arterial trees for large organs \cite{nagy2009}. Accurate modelling of
biological angiogenesis in embryo development could mimic the
development of the adult vasculature but has not yet been achieved for
large organs.

To simulate large arterial structures, a local optimisation technique,
called Constrained Constructive Optimisation (CCO), has been used
extensively to generate large arterial trees
\cite{Schreiner1993,Karch2000}. Large vasculature structures generated
using CCO are energetically sub-optimal since the optimisation is
local and there are no costs to intersecting areas of functional
tissue. Local optimisation at individual vessel junctions does not
generally result in the most efficient overall tree. CCO also has
limitations when applied to hollow organs, and tends to generate trees
that are too symmetrical (especially for the largest
arteries)\cite{Schreiner1997,Schreiner2006}.  This situation can be
improved by combining CCO with medical imaging of e.g. the large
cerebral arteries, leading to impressive results\cite{linninger},
however, there is no single algorithm based on CCO that can handle all
of the required length scales. The SAVO algorithm presented here is
capable of approaching the global cost minimum, handling complex organ
shapes and excluding large vessels from tissue. This contrasts with
CCO, which needs to be adapted for each situation and struggles to
reproduce the vasculatures of complex organ shapes. It also contrasts
to morphologically based arterial growth algorithms, which require
detailed experimental data to run, making application to new organs
difficult.

Kaimovitz {\it et al.}\cite{Kaimovitz2010} previously developed a
hybrid approach, making heavy use of morphological data to grow very
large trees featuring both arteries and veins. The trees grown are
impressively large, but their method is difficult to
generalise. Treatment at different Strahler orders varies according to
an {\it ad-hoc} scheme.  On initialisation, the branching structure of
their trees is selected randomly to follow morphological constraints,
but from that point on, the tree topology is fixed. They use simulated
annealing to optimise the orientations of the epicardial part of this
structure subject to constraints, but not to optimise the
topology. The major differences with the algorithm presented here are
that (1) swap node updates are included that are able to
explore the full configuration space of the topology of the tree,
rather than setting up the tree structure on initialisation (2) all
levels in the tree are treated with the same universal set of
principles (3) experimental data is not required as an input to this
algorithm, with the exception of the tissue shape, so any agreement
with morphological data is a direct result of the algorithm and is not
caused by the introduction of experimental morphological data into the
algorithm.

Overall, the algorithm and model presented here have significant
potential. SAVO has been shown to be capable of growing detailed
vascular trees for two large organs with complex vasculature (the
heart and brain). The {\it in-silico} model presented here matches
morphological data, and reproduces features that would be difficult, if
not impossible, to reproduce with other available algorithms, without the
need for detailed measurements of morphological data. Obtaining this level of
detail over such a large structure would also be a major challenge for
imaging techniques (e.g. the diameters of the arteries here are around
one third the diameters obtainable using MRI imaging). Further
extensions should demonstrate the possibility of growing arterial and
venous vascular structures simultaneously. Greater efficiency would
allow the growth of much larger trees (including multiscale growth),
and the possibility of describing tortuous vessels by introducing kink
nodes without bifurcations to the algorithm.

As improvements are made to the algorithm, additional applications are
expected. SAVO could be used to fill in gaps in angiography imaging in
a similar manner to Linniger {\it et al.}\cite{linninger}. The
algorithm could be used to design vasculatures for artificial
tissue. The algorithm also has immediate applications in any problem
that requires knowledge of the flows from the cerebral arteries, such
as stroke modelling\cite{hague2013}.

\section*{Acknowledgements}

JK acknowledges support from EPSRC grant EP/P505046/1. EC acknowledges
support from EPSRC grant EP/L025884/1. The authors declare no competing
interests. The authors would also like to thank Mark Horsfield (Xinapse
systems) for help with the MRI scans.

\section*{Authors' contributions}

JK developed the codes, acquired, analysed and
interpreted data. EC managed MRI data and co-supervised the
project. JPH conceived the study, developed initial versions of the
algorithm, contributed to the codes, acquired,
analysed and interpreted data, and supervised the project. All authors
contributed to writing the article.

\bibliographystyle{unsrt}
\bibliography{references}

\section*{Appendix: Modified diameter defined Strahler order scheme}

The diameter defined Strahler order (DDSO) scheme from
\cite{Jiang1985} uses radius information to improve the Strahler
order scheme. The algorithm in \cite{Jiang1985} relies on the use of
the standard deviation of the radii within individual Strahler orders,
which is only strictly valid if the data are Gaussian distributed. In
practice this is not guaranteed, and is in fact not expected since the
iterative DDSO algorithm introduces a lower radius cutoff at each
Strahler order, thus skewing the data.

A more general modified diameter defined Strahler order (MDDSO) scheme
is proposed that can handle skewed data. For convenience, the 25th and
75th percentiles are used to define the bins, although any percentiles
could be used (e.g. 15th and 85th, which approximately match the
standard deviation if data are Gaussian). Then the algorithm is as
follows:

\begin{enumerate}
\item Calculate the Strahler order.
\item Determine 25th and 75th percentiles for radii in each order, $P^{(i)}_{25}$ and $P^{(i)}_{75}$ respectively, where $i$ represents order.
\item Scanning from the end nodes of the tree, when two vessels meet, identify the vessel with largest order, $O$. This order increments in the parent vessel if and only if the radius is greater than $(P^{(O)}_{75}+P^{(O+1)}_{25})/2$
\item Repeat steps (ii)-(iv) until the mean radius at each order converges.
\end{enumerate}

\begin{figure}
  \centering
\includegraphics[width=0.48\textwidth]{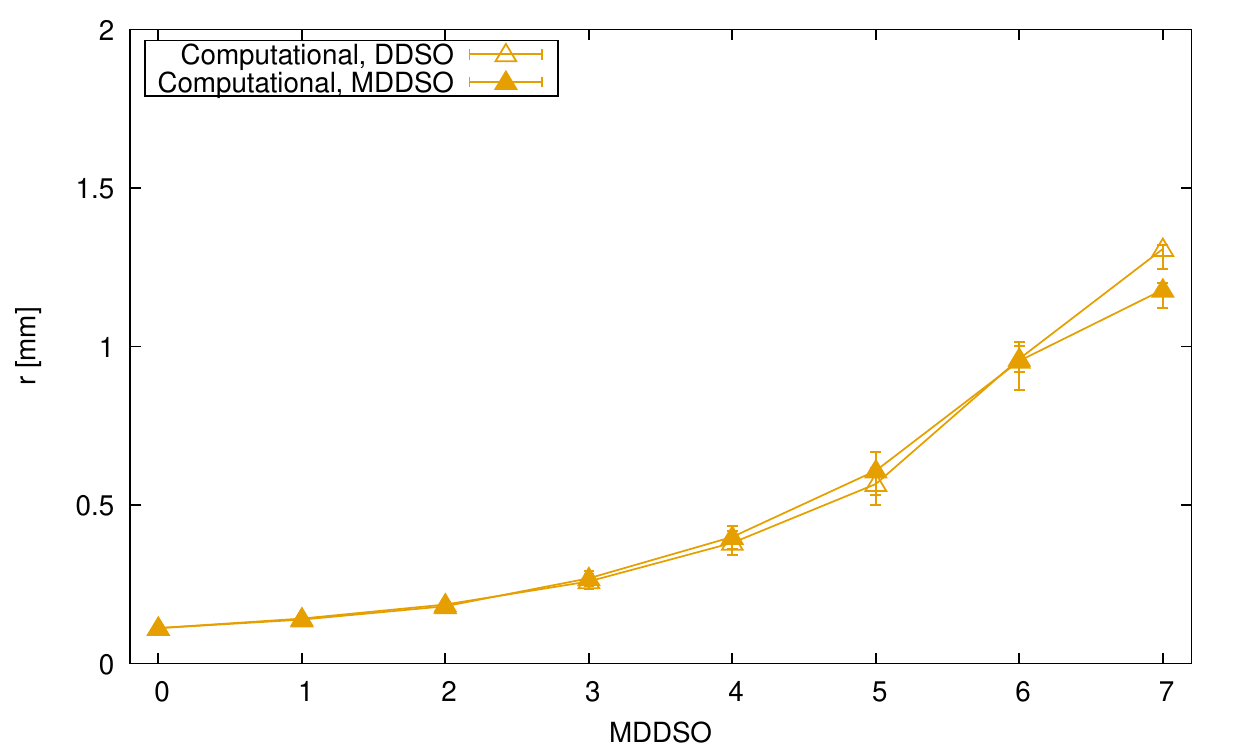}
\includegraphics[width=0.48\textwidth]{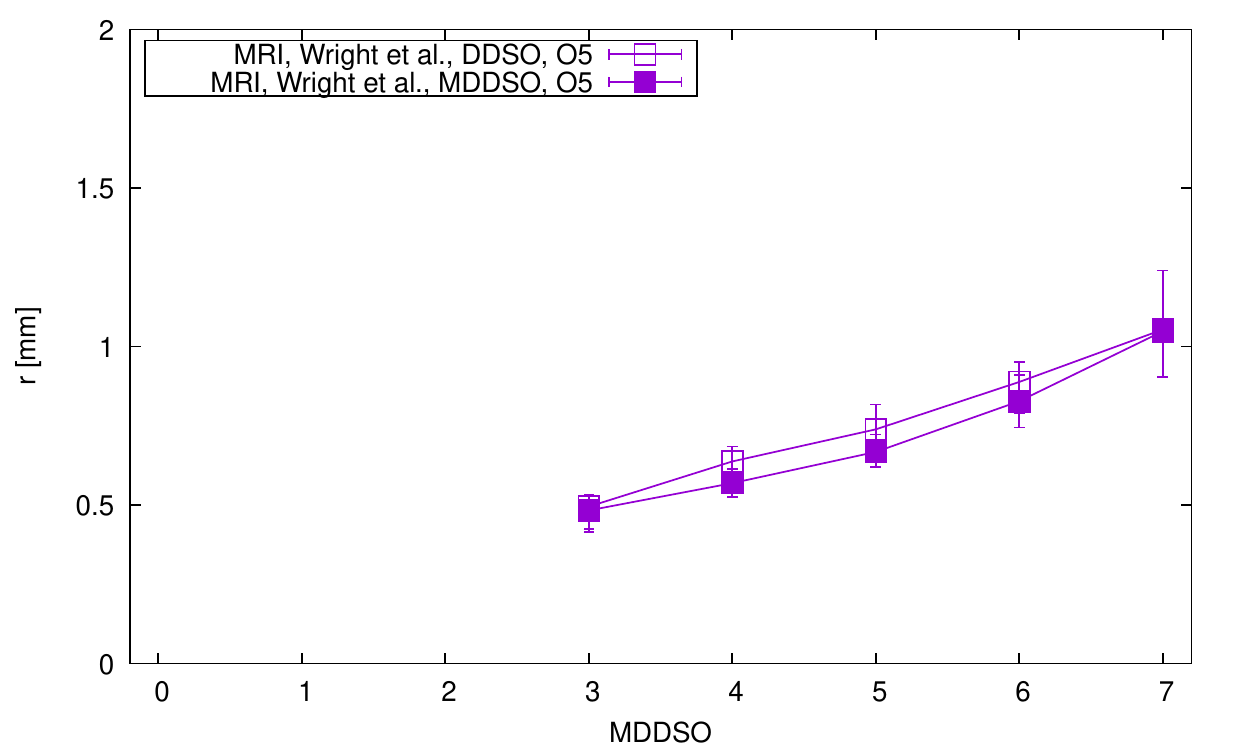}
\includegraphics[width=0.48\textwidth]{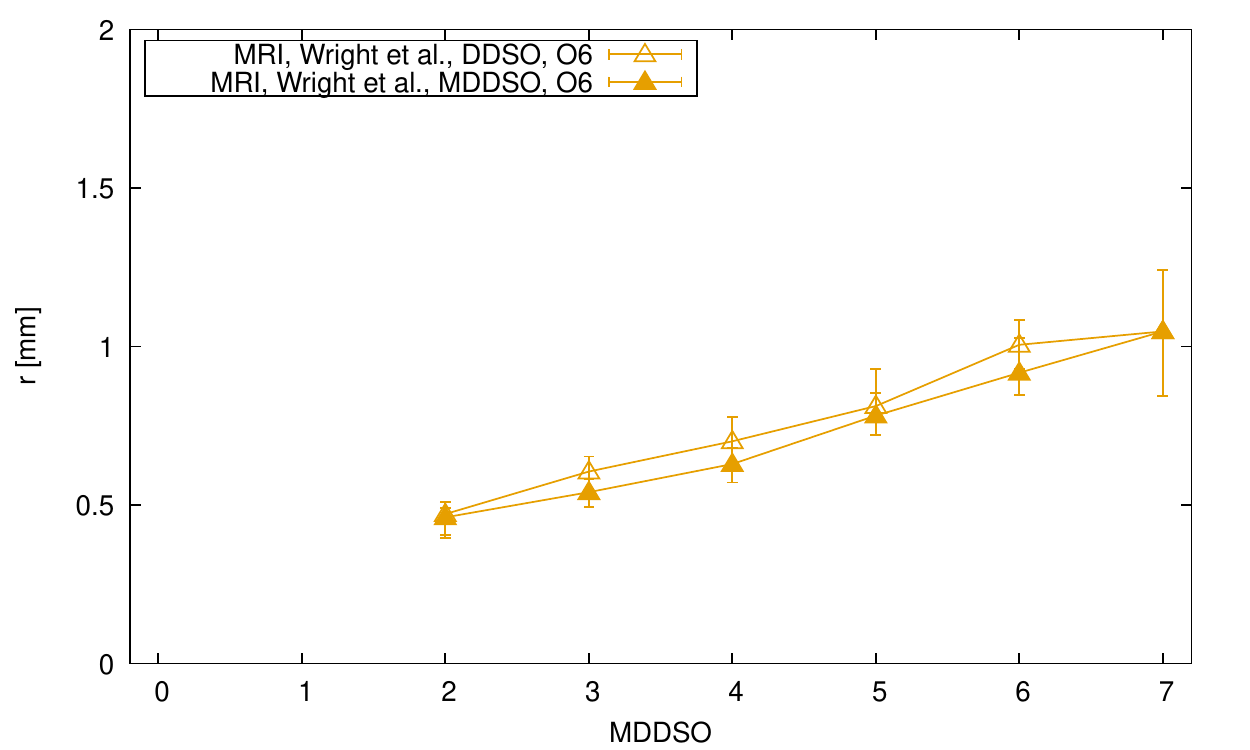}
\caption{Comparisons of MDDSO with DDSO.}
\label{fig:mddsovsddso}
\end{figure}

Examples of how results change with the modified scheme can be seen in
Fig. \ref{fig:mddsovsddso}. The changes are relatively minor, with the
main difference a small reduction of the mean radius at each order. If
15th and 85th percentiles are used, and data are Gaussian, then the
two schemes are expected to be identical.

\end{document}